\documentclass[aps,prl,reprint,amssymb,superscriptaddress,longbibliography,preprintnumbers,twocolumn,amsmath]{revtex4-1}

\usepackage{graphicx}
\usepackage{amsmath,bm}
\usepackage{hyperref}
\usepackage{physics}
\begin{document}

\title{Edge-state-induced correlation effects in two-color high-harmonic generation}

\author{Simon Vendelbo Bylling Jensen}
\affiliation{Department of Physics and Astronomy, Aarhus
University, DK-8000 Aarhus C, Denmark}

\author{Hossein Iravani}
\affiliation{Department of Chemistry, Tarbiat Modares University, P.O. Box 14115-175, Tehran, Iran}

\author{Lars Bojer Madsen}
\affiliation{Department of Physics and Astronomy, Aarhus
University, DK-8000 Aarhus C, Denmark}

\date{\today}

\begin{abstract}
We show that two-color high-harmonic spectroscopy can reveal finite-size and edge-state-induced dynamic electron correlation effects in a generic nanostructured band-gap material. 
Compared to the response of a bulk sample, we demonstrate a significant correlation-induced increase in the efficiency of the generated signal over a wide range of frequencies by harnessing the power of these ultrafast many-electron dynamics.
\end{abstract}
\maketitle

Since the demonstration of high-harmonic generation (HHG) in solids \cite{Ghimire2011}, the process has attracted increased interest due to its capability (i) to produce ultrafast coherent light pulses of high frequency \cite{Schubert2014, Luu2015a, Garg2018} and (ii) to retrive information about electron dynamics on an ultrafast timescale~\cite{Kruchinin2018}. The mechanism responsible, although still being debated \cite{Tamaya2016, Yue2020a,Yue2020b,Lakhotia2020,Ghimire2012, Hawkins2015, Hohenleutner2015a, Wu2016a, You2017, Li2019a, Ikemachi2017, Higuchi2014}, is typically explained through a combination of intraband 
\cite{Ghimire2011,Ghimire2012} and interband transitions \cite{Vampa2014}. The physics of the interband transitions has basic similarities with the three-step model of HHG in atoms \cite{Kulander1993,Corkum1993,Lewenstein1994a}. 
Most of the theoretical frameworks \cite{Higuchi2014, Hawkins2015, Ghimire2011, Ghimire2012, Vampa2014, Schubert2014, Li2019a, Luu2015a, You2017, Hohenleutner2015a, Wu2016a, Ikemachi2017, Garg2018} describing the behaviour of HHG in solids are based on a bandstructure picture. This independent-electron approach may be justified when the beyond-mean-field electron-electron interaction is of minor importance for the initial state as well as for the HHG process. We may understand the absence of electron correlation effects in HHG by considering that for the typical applied laser intensities there is only a limited excitation from the valance to the conduction band and, consequently, the density only changes slightly. The minor role of correlation in some condensed matter systems has been verified for single-pulse HHG scenarios by \textit{ab initio} approaches \cite{Tancogne-Dejean2017a,Tancogne-Dejean2017b,Hansen2017,Yu2020} and has been further verified for disordered \cite{Yu2019}, doped \cite{Yu2019a}, topological \cite{Bauer2018}, finite \cite{Hansen2018} and systems with vacancies \cite{Iravani2020}. 
Exceptions to this general trend occur in case of enhancement beyond the first plateau where correlation effects may promote specific parts of the HHG spectra, as shown in theoretical works \cite{Ikemachi2018,Yu2019} and alluded to in experiment \cite{Ndabashimiye2016}. Decrease in HHG signal is found within a wide range of HH frequencies as a result of correlational effects in two dimensional materials \cite{Tancogne-Dejeane2018,Le_Breton2018}. We note that if such correlational dynamics can be controlled and reverted, they could provide a pathway for significant enhancement across a wide range of HH frequencies. In search of utilizing and characterizing such correlational effects we thereby enter the finite-size regime of nanostructured materials, where electron-hole interactions become of higher importance for the HHG process \cite{Liu2017}, and where electron-electron effects are expected to become relevant \cite{DeVega2020}. Furthermore, for such nanostructures, even within the single-active-electron (SAE) approach, a significant increase in efficiency can be contributed to quantum confinement \cite{McDonald2017}, and if the system is sufficiently thin, the propagation damping 
can be reduced~\cite{Kilen2020}.

Recently, many-electron correlation effects of HHG have been in focus, when considering strongly-correlated effective-Hamiltonian models \cite{Imai2019, Murakami2018, Silva2018, Takayoshi2019, Tancogne-Dejean2018, Lysne2020}, such as the Fermi-Hubbard model. For such cases, HHG spectroscopy can uncover rich ultrafast highly nonequilibium many-electron dynamics that can be time-resolved through its dynamical effect on the HHG spectrum. It seems like an obvious question to consider, whether one can achieve control of these strong many-electron correlated Coulomb interactions while utilizing the efficiency of quantum confinement,  to create correlational enhancement for efficient HHG in a nanoscale system. 

In this work, we discover methods of inducing strong electron correlation effects in a two-color HHG scheme for a nanostructured band-gap material. Furthermore, we characterize the origin of these effects and demonstrate their ability to create a significant enhancement of the HHG yield over a wide spectral range.

We consider contributions from all bands in the band-structure, go beyond the tight-binding approximation and take dynamical electron-electron interaction into account applying time-dependent density-functional theory \cite{Runge1984}. No macroscopic propagation effects are accounted for. Such effects may modify the HHG spectra through absorption or interference.  Our approach is valid for thin target materials, which is feasible experimentally \cite{Liu2017,Garg2016,Nakamura2020,Taucer2017,Yoshikawa2017}, where such propagation effects are suppressed \cite{Kilen2020}. We consider nanoscale precision, which is obtainable in production of, e.g., ZnO nanowires \cite{Grinblat2014}.  
To make the discussion as general as possible, we consider a nanoscale model-sample of a generic band-gap material with band-gap energy of multiple IR photons,  as well as its bulk limit. Features of the present findings can thus be expected for a range of band-gap materials if accounting for system-specific rescaling of, e.g., laser parameters. We work in atomic units  
with a linear chain of $N$ ions, each with nuclear charge $Z = 4$ generating a static ionic potential, $v_{\mathrm{ion}} \left( x\right) = - \sum_{i=0}^{N-1} Z [\left(x-x_i \right)^2 + \epsilon]^{-1/2}$ for nuclear coordinates $x_i  = \left[ i -  \left( N - 1 \right)/2\right] a$, separated by lattice constant $a= 7$. To model the nanoscale system, we consider $N=80$, which  matches the $\sim 30$ nm diameter of ZnO nano rods \cite{Grinblat2014}. We consider a charge and spin neutral system with the 3D expression for local spin-density exchange and apply a softened Coulomb potential, with the softening parameter $\epsilon = 2.25$, to mimic 3D electrons driven in the polarization direction of linearly polarized light-field.
We have performed extensive scans over system size, and 
the reported findings are consistent in the range of $60 \leq N \leq 160$, where below $N=60$ we observe behavior of atomic nature \cite{Hansen2018} and above $N = 160$, we observe convergence to the bulk response \cite{Yu2020}. The system of choice is identical to that examined in Ref.~\cite{Hansen2017}, designed to represent a similar system as of Ref.~\cite{Wu2015}. It is ideal for comparisons with previous theoretical works as it has been studied extensively in the literature, where detailed descriptions can be found
\cite{Hansen2017, Hansen2018, Bauer2018, Yu2019a, Yu2019, Iravani2020,Yu2020}. When applying an electromagnetic field described by its vector potential, $A\left(t\right)$, the Kohn-Sham (KS) orbitals, $\varphi_{\sigma,i} \left(x ,t \right)$, are propagated through the time-dependent KS equation $i \partial_t \varphi_{\sigma,i} \left(x,t \right) = \left\lbrace - \partial^2_x/2 - i A\left(t \right) \partial_x + \tilde{v}_{\mathrm{KS}} \left[n_\sigma\right] \left(x,t \right)  \right\rbrace \varphi_{\sigma,i} \left(x ,t \right)$ using the time-dependent KS potential 
\begin{equation}
\tilde{v}_{KS}\left[ \left\lbrace n_\sigma \right\rbrace \right] \left( x ,t\right) = v_{\mathrm{ion}} \left(x\right) + v_H \left[ n \right] \left( x ,t \right) + v_{xc}\left[ \left\lbrace n_\sigma \right\rbrace \right] \left( x,t \right). \label{eq:kspot}
\end{equation}
The dynamical part hereof consists of the Hartree potential $v_H \left[ n \right] \left( x,t \right) = \int dx' n\left( x',t\right)[(x-x')^2 + \epsilon]^{-1/2}$, and the exchange-correlation potential, which in the local spin-density approximation is given as $v_{xc}\left[ \left\lbrace n_\sigma \right\rbrace \right] \left( x,t \right) \simeq  - \left[ 6 n_\sigma (x,t) \right/\pi]^{1/3}$. The spin density and the total density are given, respectively, as $ n_\sigma \left(x,t\right) = \sum_{i=0}^{N_\sigma -1} \abs{\varphi_{\sigma,i} \left( x ,t\right)}^2, $ and $ n\left(x,t\right) = \sum_{\sigma = \uparrow, \downarrow} n_\sigma \left(x,t\right)$ with $N_\sigma$ being the number of electrons with spin $\sigma =\lbrace\uparrow,\downarrow\rbrace$. We have checked that the conclusions of the present work remain unaffected by using another functional \cite{PBE1996}. The orbitals obtained by imaginary time propagation are propagated using the Crank-Nicolson method with a predictor-corrector step applying an absorbing potential \cite{Bauer2017}. In multiple earlier single-IR-pulse studies, \cite{Hansen2017,Hansen2018,Bauer2018,Yu2019a,Yu2019,Iravani2020}  all-electron dynamics were well-described in the independent many-electron band-structure picture using a "frozen" calculation, corresponding to setting $\tilde{v}_{KS}\left[ \left\lbrace n_\sigma \right\rbrace \right] \left( x ,t\right) \simeq \tilde{v}_{KS}\left[ \left\lbrace n_\sigma \right\rbrace \right] \left( x ,0 \right)$. This is equivalent to a non-interacting electron calculation, where all electrons move independently in a static, effective potential which contains the electron-electron interaction but only for the initial static system. We employ the notation of being correlated if going beyond this frozen approach during the HHG process, see also Refs.~\cite{Tancogne-Dejean2017a, Tancogne-Dejean2018,Tancogne-Dejeane2018}. By including the dynamical electron-electron interaction, the  full "dynamical" propagation method can be applied with a time-dependent KS potential $\tilde{v}_{KS}\left[ \left\lbrace n_\sigma \right\rbrace \right] \left( x ,t\right)$, and the approach captures electron correlation effects. The HHG spectra are calculated through the modulus squared of the Fourier transformed current.  
 
\begin{figure}
\includegraphics[width=0.49\textwidth]{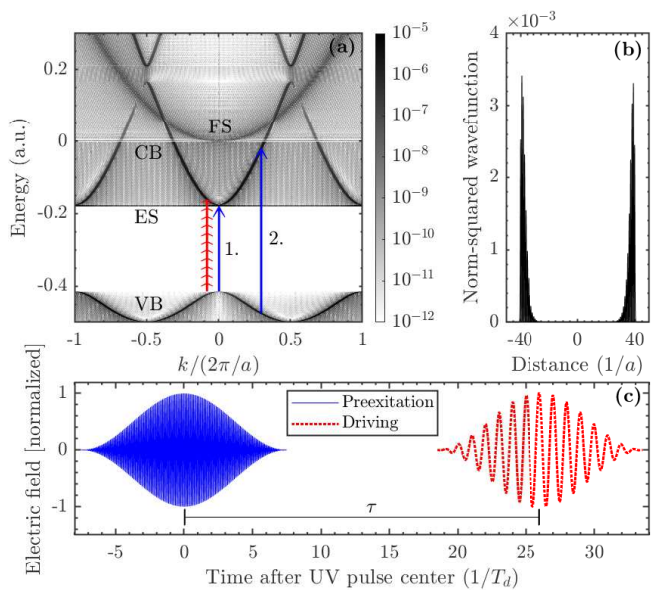}
\caption{(a) Band structure with the highest energy VB and the lowest energy CB. The ESs are indicated below the CB by the horizontal line at $\approx - 0.18$. The free-space (FS) dispersion is visible. Arrows denote central photon energies, the smallest red arrows depicting the $11 \omega_d$ multiphoton transition, 1., illustrating the preexcitation pulse when resonant with the ES, and, 2., the preexcitation pulse off-resonant with the ES. (b) The norm-squared wave function in real-space of the ES below the first CB for an $N=80$ system. (c) The pulse sequence with the UV preexcitation and the IR driving laser pulse in  units of the period of the IR laser, $T_d$, and with each pulse normalized individually. The pulses drive the transitions in (a), see text for parameters.} \label{fig:1}
\end{figure}

For large systems, a Fourier transformation of the KS orbitals resolve the band structure \cite{Hansen2018}. The band structure consists of two completely filled valence bands (VBs) and numerous unoccupied conduction bands (CBs). What is very interesting about finite-size systems including this one, is that several edge-states (ESs) appear; in this model just below each VB and CB. Here we examine the interaction with two ESs just below the lowest energy CB \cite{Yu2020}. The highest energy VB, and lowest energy CB are depicted in Fig.~\ref{fig:1}(a) for an $N=80$ system, with the ES energy shown by the horizontal line at an energy of $\approx -0.18$. One of these ESs is depicted in real-space in Fig.~\ref{fig:1}(b). Due to their localization in real-space, they are delocalized in momentum space and all $k$-values within the first brillouin zone are contained in a single ES. Such ESs have been seen to dominate the electron dynamics in low dimensional elongated systems as graphene nanoribbons and carbon nanotubes \cite{Hod2007}.

\begin{figure}
\includegraphics[width=0.49\textwidth]{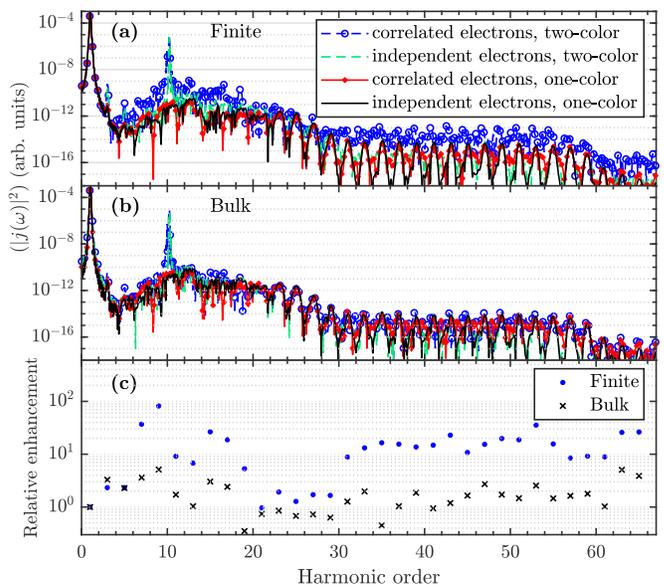} 
\caption{ HHG spectra (a) for a finite-size system ($N=80$),  and (b) for a bulk system for preexcited and IR-only ($n_d = 15$-cycle IR-pulse with $\omega_d = 0.023$ and $F_{0,d} = 0.00552$) references in correlated and independent-electron descriptions. (c) Ratio between the preexcited signal including correlation and the preexcited independent-electron signal  for $N=80$ and the bulk system.  
The preexcited systems were prepared at  a time $\tau = 26 T_d$ earlier using a $n_p = 153$-cycle UV-pulse with $\omega_p = 0.235$, $F_{0,p} =0.0005$, and resonant with the ESs [see Fig.~\ref{fig:1}(a) and (c)].  
} \label{fig:2}
\end{figure}

The applied linearly polarized electromagnetic field is illustrated in Fig.~\ref{fig:1}(c) and can be described by $
F\left(t \right) = F_{p}\left(t \right) + F_{d}\left(t \right)$, where the subscripts $p$ and $d$ denote a preexcitation and driving pulse, respectively, and can be thought of as a pump-probe setting. Both pulses are sinusoidal waves with a $\sin^2$ envelope function for the vector potential, and are for most cases long enough to neglect carrier envelope phase effects, see, e.g., Ref.~\cite{Madsen2002f}. The $15$-cycle driving IR-pulse is fixed with a frequency of $\omega_d = 0.023$ ($\lambda_d \simeq 2 \mathrm{\mu m}$). As seen in Fig.~\ref{fig:1}(a) an interband process requires at least an $11$-photon transfer. The driving field has amplitude $F_{0,d} = 0.00552$ corresponding to an intensity of $\simeq 10^{12} \mathrm{W/cm}^2$, as used in Refs.~\cite{Hansen2017,Iravani2020}. 
Applying similar parameters as in earlier two-color studies \cite{Lu2019}, we at first consider the pump-probe setting of Fig.~\ref{fig:1}(c) with similar pulse durations and a peak-to-peak delay of $\tau= 26 T_d$, which is within the spontaneous decay time of the carriers \cite{Lu2019}. Here $T_d= 2 \pi / \omega_d$ is the period of the driving frequency. We apply a $153$-cycle preexcitation pulse tuned to either resonantly couple to the ESs 
with $\omega_p = 0.235$ or to couple to the CB  
with $\omega_p = 0.455575$, which is off-resonant with the ESs [Fig.~\ref{fig:1}(a)]. The preexcitation pulse allows for controlled population of carriers, 
which is capable of selectively enhancing specific harmonics within the independent electron approach \cite{Lu2019,Song2020} and not taking ESs into account.  Opposed to previous two-color studies, we go beyond the independent-electron approach and explicitly target multielectron correlation effects to identify the characteristics hereof. To examine nanoscale-related behavior, we populate the ESs of the system with a preexcitation pulse [see Fig.~\ref{fig:1}(a), 1.~arrow], and find a significant enhancement when including electron correlation. This is shown in Fig.~\ref{fig:2}, where the parameters are given in the caption. When comparing the one-color reference system to the preexcited system, we see the characteristics of the one-photon resonant transition from the VB to the ES, visible as a peak $\omega \sim 10.2 \omega_d$, corresponding to the frequency of the preexcitation pulse $\omega_p$. We have verified by a calculation applying only the preexcitation pulse, that this peak occurs as a perturbative response of the preexcitation pulse. For the finite system a correlational enhancement appears when comparing the independent-electron and the correlated dynamical methods. For the bulk system, no such correlation effects appear, solidifying the finite-size characteristic of this effect. The bulk response is found by an $N=300$ calculation, in which the spectrum has converged to the spectra of a periodic system~\cite{Yu2020}.  
As is seen by Fig.~\ref{fig:2}(c), the correlation enhancement is present at a wide range of frequencies, reaching into the first plateau entering the range of most current experimental detection. 

\begin{figure}
\includegraphics[width=0.49\textwidth]{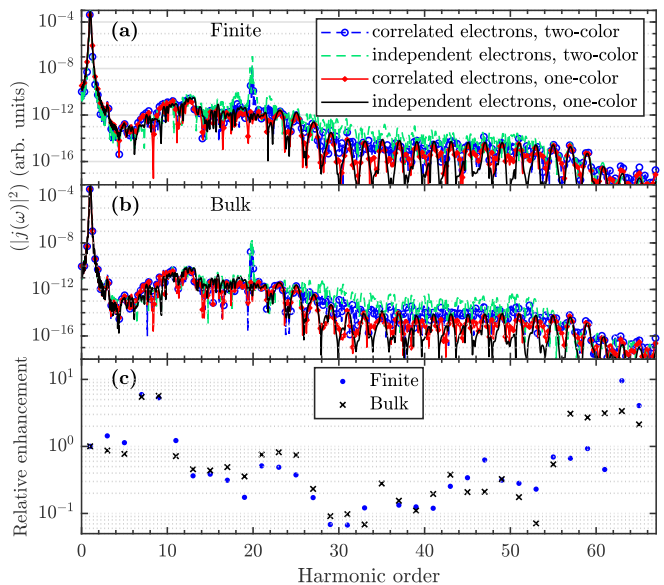}
\caption{As Fig.~\ref{fig:2}, but for $\omega_p = 0.455575$, off-resonant with the ESs [see Fig~\ref{fig:1}(a)]} \label{fig:3}
\end{figure}
The origin of this correlational enhancement relies in populating the ES. Since this state is localized in real space, and de-localized in momentum space, transitions hereto are available for every state in the VBs, at a wide range of frequencies. To highlight the role of the resonantly populated ES, a  non-resonant pump-probe calculation with a frequency of $\omega_p = 0.455575$ was made [see Fig.~\ref{fig:1}(a) 2. arrow]. Applying this preexcitation frequency there are no resonant transitions available to the ES within the spectral range. The resulting spectra in Fig.~\ref{fig:3} show no finite-size correlational enhancement, as the finite-system response coincides relatively well with the bulk response. The characteristics of the one-photon resonant transition from the VB to the CB is visible as a peak at $\omega \sim 19.8 \omega_d$, corresponding to the frequency of the preexcitation pulse $\omega_p$. Our study indicates that a two-color scheme is able to promote finite-size effects and increases the limit for bulk-dominated dynamics from $N \sim 60$ ($\sim 22$ nm for the present lattice constant) seen in Ref.~\cite{Hansen2018} to $N \sim 220$ ($\sim 81$ nm), even further than the converged one-pulse results discussed in Ref.~\cite{Yu2020} further elucidating the range of sizes where the ESs dominate the electronic behaviour \cite{Hod2007}.
\begin{figure}
\includegraphics[width=0.49\textwidth]{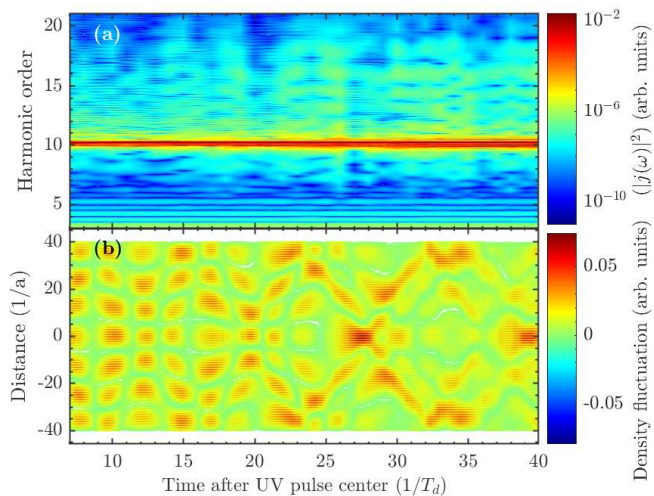}
\caption{(a) HHG spectra of a finite ($N=80$) system as a function of time-delay between preexcitation by an $n_{p}= 153$-cycle UV-pulse with $\omega_p = 0.235$ and $F_{0,p} =0.0005$ and a $n_{d}= 2$-cycle IR-pulse with $\omega_d = 0.023$ and $F_{0,d} = 0.00552$ driving pulse, applying correlated dynamical propagation. (b) The density fluctuations $n(x,t)-n(x,0)$ of the system applying the correlated dynamical method given as a function of time, with parameters of (a) but with no driving field.} \label{fig:4}
\end{figure}

In order to characterize the electron dynamics underlying the correlation enhancement, a time-resolved investigation is useful. This is done with the correlated propagation method applying various peak-to-peak delays, $\tau$, [Fig \ref{fig:1} (c)]. To resolve the dynamics in time, a $2$-cycle driving pulse with a vanishing carrier envelope phase was applied. The resulting spectra are depicted in Fig.~\ref{fig:4}(a) as a function of $\tau$. This figure shows that the HHG spectra depends strongly on $\tau$ and hence on the electron dynamics in the nanoscale material after the preexcitation pulse. A direct way of gaining insight into the underlying dynamics, is by considering the time-dependent density of a preexcited system as in Fig.~\ref{fig:4}(b). By resonant coupling of the ESs, the edge of the system is populated during the preexcitation pulse. Soon hereafter the populated localized ESs will create an increased repulsion, resulting in an electron wave propagating from the edges inwards towards the center of the sample. The dynamical ES-induced electron-density fluctuation propagates and interferes throughout the system [Fig.~\ref{fig:4}(b)]. This fluctuation of electron density leads to a time-dependent change in the KS potential, which breaks the translational symmetry within the sample. The density fluctuates after the preexcitation pulse has ended at $7.5 T_d$ even though nothing is driving the system.  
The dynamical KS potential, dominated by the Hartree term, will drive the system away from regions, where the interfering density fluctuation is localized. This intrinsic time dependence, initialized by the preexcitation, may continue to drive electrons and further induce currents to produce harmonics after the end of the preexcitation pulse. When applying a driving pulse, its interaction with the time-dependent electron density will be $\tau$-dependent, and result in $\tau$-dependent spectra. Such correlation effects were revealed in Fig.~\ref{fig:2} to contribute with an important effect when applying a driving field to a preexcited nanoscale system. One can observe directly in Fig.~\ref{fig:4}, that the time-dependent changes of the density, are revealed through changes of the HHG spectra. Thus two-color HHG spectroscopy provides a pathway to resolve coherent  many-electron dynamics in finite systems. We have found that this delay-dependency is also present for the $15$-cycle driving field of Fig.~\ref{fig:2}, even though the timescale of the fluctuations cannot be resolved. This indicates that by tuning $\tau$ of a pump-probe HHG scheme, one can experimentally test for indications of dynamics beyond independent-electron approaches. The calculations furthermore suggest that the ability to tune $\tau$, which is a simple experimental procedure, is crucial for achieving optimal correlational enhancement. We note that modelling of the two-color scheme of Fig.~4 by an independent-electron simulation (with frozen KS potential), will show no variation of the density as a function of $\tau$ after the end of the preexcitation pulse, as no repulsion is initialized in the ESs. For such a calculation a slight $\tau$ dependency can still be witnessed in the HHG spectra (not shown), as the wavepacket initialized by the preexitation pulse is probed at different times by the driving pulse. This $\tau$ dependency will, however, in a full simulation be completely overshadowed by the large correlation effects, from which the $\tau$ dependency of Fig.~4 is attributed.

In conclusion, the present results elucidate the importance of considering electron correlation effects, when investigating HHG in nanosize band-gap materials. Furthermore, we characterize the correlational enhancement as a consequence of ES-induced electron fluctuations, and show that these many-electron dynamics can be revealed through HHG spectroscopy, or harnessed for enhancement in two-color HHG, for a wide range of harmonic frequencies. As the localized ESs initialize the dynamics, it is interesting to investigate in the future if similar effects can stem from localized atom-like orbitals introduced by, e.g., doping.
\begin{acknowledgements}
This work was supported by Danish Council for Independent Research(GrantNo.9040-00001B).
\end{acknowledgements}

\end{document}